\documentclass[aps,prl,twocolumn,superscriptaddress]{revtex4}

\usepackage{graphicx,amsmath}
\usepackage{color}

\usepackage{booktabs}
\usepackage[flushleft]{threeparttable}

\usepackage{amssymb}
\usepackage{amsmath}
\usepackage{float}
\usepackage{psfrag}
\usepackage{siunitx}
\usepackage{ulem}

\def\be{\begin{equation}}
\def\ee{\end{equation}}
\def\ba{\begin{eqnarray}}
\def\ea{\end{eqnarray}}

\begin{document}

\title{Gravitational effect in evaporating binary microdroplets}
\author{Yaxing Li}
\affiliation{Physics of Fluids group, Department of Science and Technology, Mesa+ Institute, Max Planck Center for Complex Fluid Dynamics and 
J. M. Burgers Centre for Fluid Dynamics, University of Twente, P.O. Box 217, 7500 AE Enschede, The Netherlands}
\author{Christian Diddens}
\affiliation{Physics of Fluids group, Department of Science and Technology, Mesa+ Institute, Max Planck Center for Complex Fluid Dynamics and 
J. M. Burgers Centre for Fluid Dynamics, University of Twente, P.O. Box 217, 7500 AE Enschede, The Netherlands}
\affiliation{Department of Mechanical Engineering, Eindhoven University of Technology, P.O. Box 513, 5600 MB Eindhoven, The Netherlands}
\author{Pengyu Lv}
\affiliation{Physics of Fluids group, Department of Science and Technology, Mesa+ Institute, Max Planck Center for Complex Fluid Dynamics and 
J. M. Burgers Centre for Fluid Dynamics, University of Twente, P.O. Box 217, 7500 AE Enschede, The Netherlands}
\author{Herman Wijshoff}
\affiliation{Department of Mechanical Engineering, Eindhoven University of Technology, P.O. Box 513, 5600 MB Eindhoven, The Netherlands}
\affiliation{Oc\'{e} Technologies B.V., P.O. Box 101, 5900 MA Venlo, The Netherlands}
\author{Michel Versluis}
\affiliation{Physics of Fluids group, Department of Science and Technology, Mesa+ Institute, Max Planck Center for Complex Fluid Dynamics and 
J. M. Burgers Centre for Fluid Dynamics, University of Twente, P.O. Box 217, 7500 AE Enschede, The Netherlands}
\author{Detlef Lohse}
\email{d.lohse@utwente.nl}
\affiliation{Physics of Fluids group, Department of Science and Technology, Mesa+ Institute, Max Planck Center for Complex Fluid Dynamics and 
J. M. Burgers Centre for Fluid Dynamics, University of Twente, P.O. Box 217, 7500 AE Enschede, The Netherlands}
\affiliation{Max Planck Institute for Dynamics and Self-Organization, 37077 G\"ottingen, Germany}

\begin{abstract} 

The flow in an evaporating glycerol-water binary sub-millimeter droplet with Bond number Bo $\ll$ 1 is studied both experimentally and numerically. First, we measure the flow fields near the substrate by micro-PIV for both sessile and pendant droplets during the evaporation process, which surprisingly show opposite radial flow directions -- inward and outward, respectively. This observation clearly reveals that in spite of the small droplet size, gravitational effects play a crucial role in controlling the flow fields in the evaporating droplets. We theoretically analyze that this gravity-driven effect is triggered by the lower volatility of glycerol which leads to a preferential evaporation of water then the local concentration difference of the two components leads to a density gradient that drives the convective flow. We show that the Archimedes number Ar is the nondimensional control parameter for the occurrence of the gravitational effects. We confirm our hypothesis by experimentally comparing two evaporating microdroplet systems, namely a glycerol-water droplet and a 1,2-propanediol-water droplet. We obtain different Ar, larger or smaller than a unit by varying a series of droplet heights, which corresponds to cases with or without gravitational effects, respectively. Finally, we simulate the process numerically, finding good agreement with the experimental results and again confirming our interpretation.

\end{abstract}


\maketitle


The evaporation of a microdroplet on a flat substrate has attracted a lot of attention because of its beautiful and phenomenologically rich fluid dynamics \cite{picknett1977,deegan1997,lohse2015rmp,hu2002,popov2005,cazabat2010,sbonn2006,ristenpart2007,schoenfeld2008,gelderblom2011,marin2011,ledesma2014,Tan2016,Gatapova2018} and its relevance in various technological applications, such as medical diagnostics \cite{brutin2011} and the fabrication of electronic devices \cite{lim2008}. For many of these applications, an understanding of the internal flow structure is crucial. One example is the so-called ``coffee stain problem"~\cite{deegan1997}, i.e. an evaporating colloidal drop in which an outward capillary flow along the substrate carries the dispersed material from the interior towards the pinned contact line. This seminal study opened up a new line of research for surface coatings and patterning technologies, which is crucial for various applications in inkjet printing~\cite{park2006control}, 3D printing technology~\cite{kong2014} and molecular biology~\cite{jing1998automated}. 

However, in nearly all of these applications, the droplet liquid is not pure, but a binary or even ternary liquid.
As is well known, then  Marangoni flow, which is  driven by surface tension gradients,  is coming into play \cite{hu2006,tam2009,christy2011,bennacer_sefiane_2014,cazabat2010}, strongly affecting the evaporative
behavior. The variation of the surface tension originates from two mechanisms or the combination of both,  namely a temperature gradient~\cite{hu2006,tam2009} or a solute concentration gradient~\cite{christy2011,bennacer_sefiane_2014,Tan2016,Tan2017,Li2018,Kim2018}, due to the spatially varying local evaporation rates at the droplet surface.
The conventional understanding is that the flows within sub-millimeter droplets can only be attributed to capillary and Marangoni convections, while natural convection is considered to be negligible as the surface tension force is dominant compared to gravity forces~\cite{tam2009}. The Bond number of such a small-sized droplet system reads $\text{Bo} = \rho g R^2/\gamma \ll 1$, which normally implies a gravity-independent system~\cite{kim2016controlled}. However, this only holds as a measure of the importance of the gravity force compared to the surface tension while relating to the droplet shape~\cite{SHIN2009}. In recent years, studies of evaporating aqueous NaCl droplets revealed that {\it natural convection} can be driven by evaporation-induced {\it density gradients} inside a colloidal droplet~\cite{SAVINO1996,kang2013,kumar2016,kumar2017,KUMAR2018}. However, to our best knowledge, an internal flow controlled by gravitational effects has never been observed, nor experimentally confirmed in a drying microdroplet consisting of two miscible liquid components.

In this work, we investigate the flow field inside an evaporating glycerol-water binary miscible droplet. Glycerol is a very common liquid heavily used in industry~\cite{soap1990} and laboratory experiments~\cite{Sousa1995}, in particular due to its strong hygroscopic nature~\cite{gly1963}. Recently, Shin \textit{et al.}~\cite{shin2016} found a spontaneous B\'{e}nard-Marangoni (BM) convection within the water-glycerol system: the hygroscopic absorption of water vapor at the interface generates thermal and solutal gradients, leading to a surface tension gradient, thus sustaining a BM instability. They confirmed that buoyancy is not the driving force by observing the same convection cells when orienting the system upside-down. Here, however, we study the reverse process to absorption: the evaporation of a dilute aqueous glycerol droplet. In this system, glycerol can be considered nearly non-volatile under room temperature as compared to water, implying a selective evaporation of the more volatile water, which leads to a concentration gradient of the components in the drying droplet. Such a concentration gradient generates surface tension and density gradients, which drives a convective flow in the droplet. Remarkably, the flow field is mainly controlled by the buoyancy force through the density gradient. 
 
\begin{figure*}[t]
\centering
\includegraphics[width=0.98\textwidth]{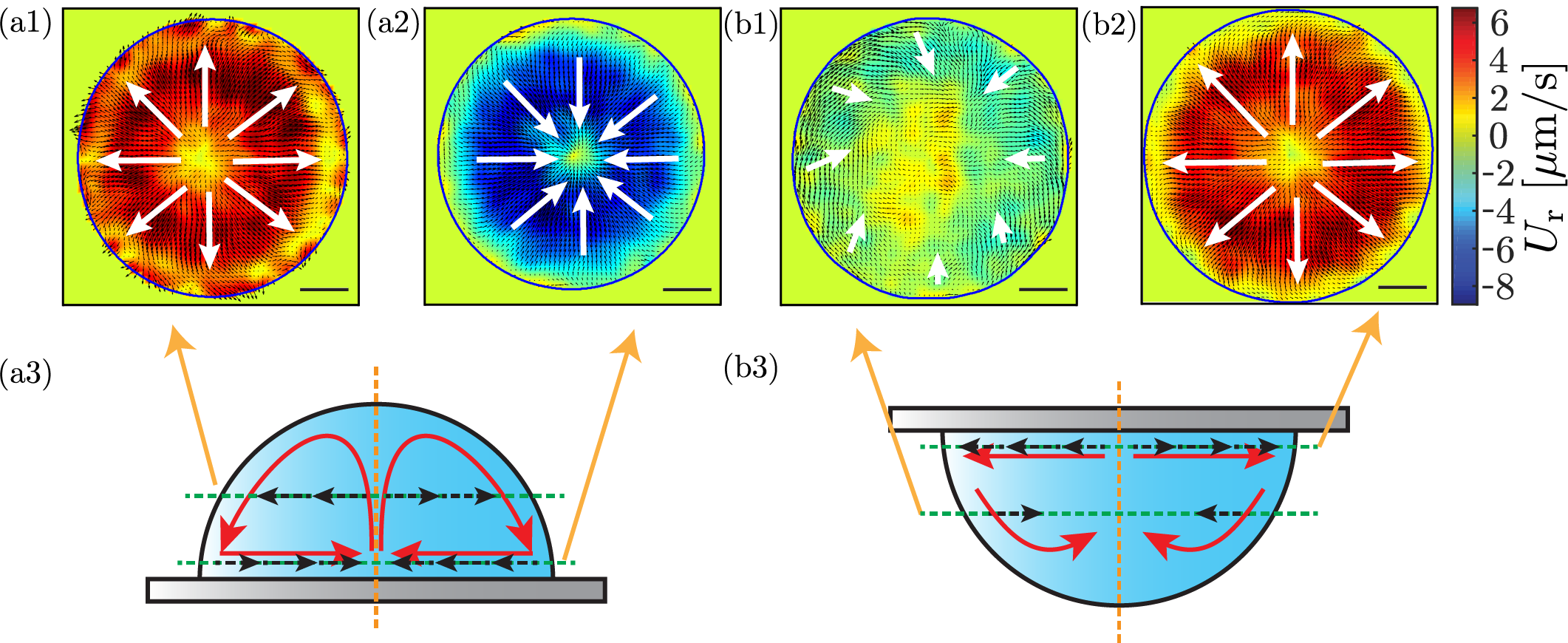}
\caption{The micro-PIV measurement of flow fields in both sessile (a) and pendant (b) droplets. The scale bars represent \SI{200}{\micro\meter}. (a1 and a2) Flow fields in a sessile binary droplet measured at different heights: \SI{200}{\micro\meter} and \SI{10}{\micro\meter} away from the substrate, respectively. The white arrows represent the flow direction. (a2) The measurement near the substrate shows an inward radial flow, (a2) the one at larger height reveals an outward radial flow. (b1 and b2) Flow fields of a pendant binary droplet measured by the same method as the sessile droplets. (b2) The flow near the substrate follows outward radial direction, (b1) but the flow at midheight reveals an annular flow with deviations from axisymmetry near the edge and irregular flow within the inner part. The four PIV images were taken with four different droplets. (a3 and b3) The schematics of the flow pattern in side views of both sessile and pendant droplets.
 }
\label{fig:exp_velocity}
\end{figure*}

We implemented microparticle image velocimetry ($\mu$PIV) to measure the flow fields within the binary droplets with opposite configuration: a sessile and a pendant droplet. We deposited (0.18 $\pm$ 0.03) \SI{}{\micro\liter} binary droplets (glycerol:water, 50:50$\%$ by weight), seeded with fluorescent microspheres (\SI{520}{\nano\meter} diameter) at a concentration of \SI{2e-2} vol$\%$ on a OTS glass substrate (see Supplemental Materials~\cite{SM}). Glycerol has a higher density $\rho_\text{g} = \SI{1.261e3}{\kilogram\per\meter^3}$ and lower surface tension $\gamma_{\text{g}} = \SI{64e-3}{\newton\per\meter}$ than water ($\rho_\text{w} = \SI{0.998e3}{\kilogram\per\meter^3}$, $\gamma_{\text{w}} = \SI{75e-3}{\newton\per\meter}$). During evaporation, the ambient temperature and relative humidity were stable, i.e., $T = 21  \pm \SI{1}{\celsius}$  and RH = 50 $\pm$ 5 $\%$. During the drying process the contact angles $\theta$ for both the sessile and the pendant droplet were between 90$^\circ$ to 104$^\circ$. A typical droplet initially has a footprint radius $R_0 = \SI{400}{\micro\meter}$ and a height $h_0 = \SI{490}{\micro\meter}$. We investigate the flow fields for both sessile and pendant droplets by adjusting the optical focus plane at different heights within the droplets:  at \SI{10}{\micro\meter} (i) and at \SI{200}{\micro\meter} (ii) away from the substrate, as labeled in Fig~\ref{fig:exp_velocity}. For sessile droplets, we observe an inward radial flow in the middle of the droplet (a1) and an outward radial flow close to the base of droplet (a2). Fig~\ref{fig:exp_velocity}.(a3) presents the schematics of the flow velocity within the axis plane of the droplet from side view: the black arrows represent the measured flow velocities and the red arrows indicate the flow pattern reconstructed from the measurement. In contrast, Fig~\ref{fig:exp_velocity}.(b1,b2) show a completely reversed flow fields for pendant droplets. The flow velocity near the substrate (b2) is radially outward, from center towards the contact line, which is opposite to the one of sessile droplet (b2). In the middle of the droplet (b1), it shows relatively weaker inward radial flow only in the outer region and asymmetric annulus flow near the edge of the droplet. Fig~\ref{fig:exp_velocity}.(b3) illustrates the flow pattern in a pendant droplet.

The opposite radial flow directions near the substrate of differently orientated droplets clearly indicates that the gravitational effect is dominating to control the flow structure. In our system, the gravitational effect is driven by the density gradient in the bulk of the droplet, which in turn is generated from concentration gradients induced by the selective evaporation of the more volatile water. As the density of the glycerol-water mixture monotonously increases with increasing glycerol concentration, the  local density decreases from the outer layer towards the inner bulk. For the sessile droplet, the denser glycerol-rich part is collecting on the top of the droplet, resulting in an unstable situation: denser glycerol-rich part suspends atop lighter water-rich part. As sketched in Fig~\ref{fig:exp_velocity}.(a3), the lighter liquid part rises up due to the buoyancy, pushing the denser liquid to sink along the outer layer, hence an inward flow is passively driven from the contact line towards the center in the bottom layer. In contrast, for pendant droplets, the glycerol-rich part is at the bottom of the liquid bulk which cannot drive a buoyancy flow in the center. Instead, the denser liquid near the contact line flows down along the outer layer, and is replenished by the outward radial flow close to the substrate, implying a much weaker gravitational effect in the pendant droplet. Moreover, due to the intense coupling between composition and flow, instabilities can arise which lead to axial symmetry breaking near the interface~\cite{christy2011,Diddens2017b}.

We can estimate the flow velocity in the evaporating binary droplet by scaling arguments. The typical velocity from the $\mu$PIV results is $U \approx \SI{10}{\micro\meter\per\second}$ so that the Reynolds number $\text{Re} = \rho UR/\mu_\text{m} \approx 10^{-4}$, where $\mu_\text{m} \approx \SI{5}{\milli\pascal\second}$ is the viscosity of the binary mixture. Furthermore, the P\'{e}clet number $\text{Pe} = UR/D \approx 10$, with the mutual diffusion coefficient $D \approx \SI{0.4e-9}{\meter^2\per\second}$, implying advection is dominating over diffusion. Assuming a quasi-steady flow, the gravitational force due to the density difference is balanced by the viscous shear stress, which scales as $g \Delta \rho_\text{m} \sim \mu \nabla ^2 U$. The density $\rho_\text{m}$ of the mixture liquid varies with the relative concentration $\phi$ following the Boussinesq approximation, $\rho_\text{m} = \rho_0[1+\beta(\phi-\phi_0)-\alpha(T-T_0)]$~\cite{nan2017}, where $\rho_0$, $\phi_0$ and $T_0$ denote, the density, concentration ratio and temperature of the reference state, respectively, and where $\alpha$ and $\beta$ indicate the thermal and solutal expansion coefficient, respectively. Note that the evaporation of water cools down the liquid temperature near the surface, which enhances the density difference between the surface area and the bulk. For now we neglected the thermal effect and only considered the density changes due to concentration differences. This assumption is validated by numerical simulations (see below) with and without the thermal contribution with only minor changes to the flow field. Thus the density difference is given by $\Delta \rho_\text{m} = \rho_0\beta\Delta \phi$, with the solutal expansion coefficient is $\beta \approx 0.2$. 

\begin{figure}[h]
\centering
\includegraphics[width=0.48\textwidth]{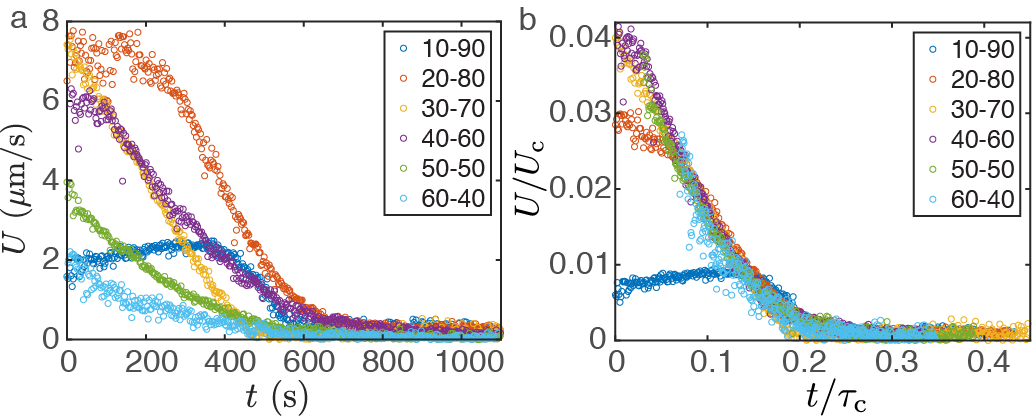}
\caption{(a) The evolutions of the averaged radial velocity near the substrate for sessile droplets with various initial glycerol mass concentrations, 10\% (blue), 20\% (red), 30\% (orange), 40\% (purple), 50\% (green) and 60\% (cyan). (b) The dimensionless averaged radial velocity plotted against the dimensionless time for the same data as in (a), which are scaled by the characteristic velocity and life time according to Eq.~\ref{eq:velocity} and Eq.~\ref{eq:lifetime} , respectively. All the scaled data collapse into each other except the early stages of the ones with 10\% and 20\% initial concentration. 
 }
\label{fig:normalized_sessile}
\end{figure}

The evaporation rate is controlled by diffusion of water vapor molecules to the surrounding air. Therefore, the mass loss $\Delta m$ within a typical timescale $\Delta t\ \sim R/U$ can be estimated from $\Delta m \sim D_{\text{w,air}} \Delta C_{\text{w,air}} R \Delta t$, with $D_{\text{w,air}} \approx \SI{e-5}{\meter^2\per\second}$ is the diffusion coefficient of water vapor at room temperature and $\Delta C_{\text{w,air}} \approx \SI{e-2}{\kilogram\per\meter^3}$ is the water vapor concentration difference between the air-liquid interface and the surrounding air. Hence, the concentration ratio difference $\Delta \phi$ can be calculated from $\Delta \phi \sim \Delta m/m_0 \sim (D_{\text{w,air}} \Delta C_{\text{w,air}}R \Delta t)/(2\pi R^3 \rho_\text{0}/3)$. Therefore, we obtain the characteristic steady-state flow velocity $U_\text{c}$ as 
\begin{equation}
U_\text{c} \sim \left(\frac{3D_{\text{w,air}} \Delta C_{\text{w,air}} R}{2 \pi \mu_\text{m}}g\beta\right)^{1/2}\ \sim \SI{e-5}{\meter\per\second},
\label{eq:velocity}
\end{equation}
thus in very good agreement with the velocity estimate from experiment. Note that Eq.~\ref{eq:velocity} is independent of the ratio of nonvolatile glycerol to water. We test the applicability of the scaling approach on various initial mass concentration of glycerol, ranging from 10\% to 60\%. The measured temporal evolutions of the averaged radial velocity near the substrate for sessile droplets are shown in Fig~\ref{fig:normalized_sessile}(a). We then
rescale the experimental data by the characteristic velocity $U_\text{c}$ and time scale $\tau_\text{c}$. The latter is estimated by the diffusion lifetime of a sessile droplet ~\cite{gelderblom2011}, 
\begin{equation}
\tau_\text{c} = \frac{\rho_\text{m}R^2}{D_{\text{w,air}} \Delta C_{\text{w,air}}}.
\label{eq:lifetime}
\end{equation}
As shown in Fig~\ref{fig:normalized_sessile}(b), with the rescaling 
all  data collapse, except the early stages of the droplets with 10\% and 20\% initial concentration. We conclude that the gravitational effect is enhanced by increasing the concentration of glycerol,  before reaching the concentration at which glycerol starts to absorb the humidity from the air.

We now introduce the Archimedes number Ar $= gh^3 \rho_0 \Delta \rho/\mu^2_\text{m}$~\cite{yu2015} as the control parameter for this problem, where $h$ is the height of droplets. For large $\text{Ar} \gg 1$ gravity plays a prominent role, whereas for small $\text{Ar} \ll 1$, the gravity can be neglected. To test the applicability of Ar, we varied the height of a series of sessile glycerol-water droplets, with $h_\text{g} = \SI{608}{\micro\meter}, \SI{514}{\micro\meter}, \SI{320}{\micro\meter}$ and $\SI{154}{\micro\meter}$. For our system, with a density difference between glycerol and water $\Delta \rho_\text{g} = \SI{263}{\kilogram \per \meter^3}$, we obtain Ar = 23.1, 14.0, 3.4 and 0.37. A prominent inward radial flow in the center of the droplets is observed for the cases when Ar is greater than 1, which indicates gravity-dominating flow. However, for the case when Ar $<$ 1, this flow disappears. For comparison, we investigated the occurrence of gravitational effects in a 1,2-propanediol-water binary droplet in the same experimental setup. The density for 1,2-propanediol is \SI{1.036e3}{\kilogram \per \meter^3} at room temperature, which is slightly greater than that of water (density difference $\Delta \rho_\text{p} = \SI{36}{\kilogram \per \meter^3}$) and $\mu_\text{p} \approx$ \SI{7}{\milli\pascal\second} for the viscosity~\cite{Tanaka1988}. As with glycerol, 1,2-propanediol is also nearly non-volatile at room temperature conditions, resulting in a preferential evaporation of water. The contact angles $\theta_\text{p}$ for sessile and pendant droplets are between \SI{70}{\degree} and \SI{75}{\degree} due to the low surface tension of 1,2-propanediol~\cite{Nakanishi1971}, indicating a stronger Marangoni effect than that of the glycerol-water droplets. We measured the flow field in an optical focal plane \SI{10}{\micro \meter} above the substrate for oppositely configured droplets (sessile and pendant) with two different droplet heights, $h_\text{p} \approx \SI{800}{\micro\meter}$ and $\SI{410}{\micro\meter}$, with Ar $\approx$ 3.7 and 0.5, respectively. We observed the opposite radial outward flows in large droplets with different orientations, similar to glycerol-water droplets. But for small droplets, we find the same radially outwards flow direction for both droplets: there is no detecable flow within the center of the droplet, but only outward radial flow in the outer regime, which clearly indicates that for these droplets with low Ar, only the Marangoni effect dominates the flow near the edge. More details are given in the supplemental materials~\cite{SM}. The two liquid systems clearly show that the Archimedes number is indeed the crucial control parameter for the occurrence of the gravitational effects.

\begin{figure}[h]
\centering
\includegraphics[width=0.48\textwidth]{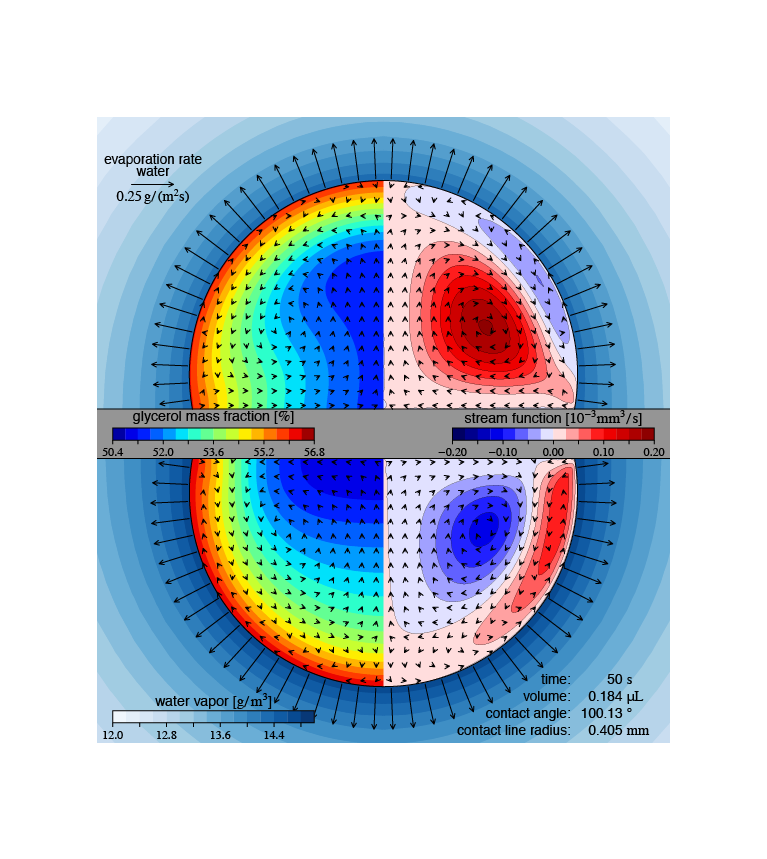}
\caption{Snapshot of the numerical results for sessile (upper) and pendant (lower) droplets, respectively. In the gas phase, the water vapor concentration (color-coded) is shown along with the evaporation rate (arrows). Inside the droplet, the glycerol mass fraction (left) and the Stokes stream function (right) are depicted, where the arrows indicate the local flow direction.
 }
\label{fig:numerical_snapshot}
\end{figure}

To validate the experimental results, corresponding finite element simulations were performed. To that end, an axisymmetric model \cite{Diddens2017b} was employed, which had already been successfully validated with binary water-ethanol and ternary ouzo droplets \cite{Diddens2017c}. The model solves the diffusion equation for the water vapor concentration $C_{\text{w,air}}$ in the gas phase, assuming vapor-liquid equilibrium according to Raoult's law. It includes the activity coefficient~\cite{marcolli2005} of the two components at the liquid-gas interface and considers the ambient vapor concentration far away from the droplet. The diffusive water vapor flux at the interface determines the volume evolution and the normal component of the velocity in the droplet and is furthermore used as interfacial sink term in the convection-diffusion equation for the local composition in the droplet. The liquid properties, including the surface tension $\gamma$, mass density $\rho$, viscosity $\mu$, diffusivity $D$ and thermodynamic activity of water $a_\text{w}$, are not constant, but coupled to the composition field. The composition-dependent properties of the glycerol-water mixture have been extracted from experimental data and are plotted in Refs. \cite{Diddens2017a,Diddens2017b}. While the gravitational body force has been neglected in the original model \cite{Diddens2017b}, the discussed experimental results clearly indicate the relevance of this term. Hence, the generalized model used here solves the Navier-Stokes flow inside the droplet with the composition-dependent body force  $\rho \vec{g}$ in the bulk and the Marangoni shear stress at the liquid-gas interface.

\begin{figure}[h]
\centering
\includegraphics[width=0.48\textwidth]{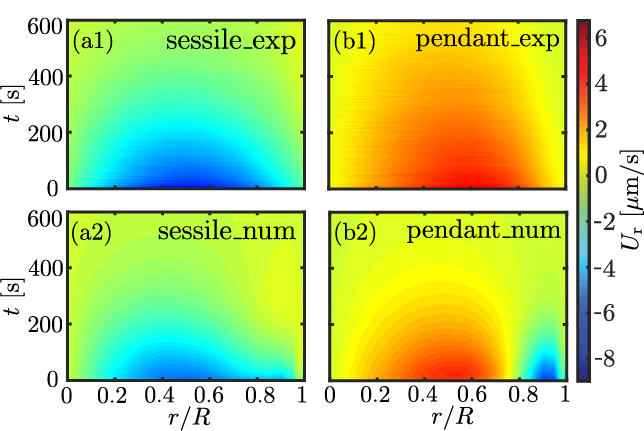}
\caption{Experimental (a1, b1) and numerical (a2, b2) results for the evolution of the radial velocity near the substrate for both sessile (a1, a2) and pendant (b1, b2) droplets. Positive values indicates outward flow. For sessile droplets (a1, a2), the numerical simulation shows a great agreement with experiment. However, for pendant droplet (b1, b2), the numerical simulation shows a stronger flow near the contact line than in the experiment, which is presumably due to the axial symmetry breaking of the flow in experiment. 
 }
\label{fig:velocity_evolution}
\end{figure}

Figure~\ref{fig:numerical_snapshot} shows snapshots of the simulations for
the sessile and the pendant droplet using the experimental
parameters. It is apparent that, as in the micro-
PIV results, close to the substrate the flow is directed inward/outward
for the sessile/pendant droplet, whereas it is reversed in the bulk
layer approximately at the center of the droplet. Along the liquid-gas
interface the simulations reveal a counter-rotating vortex as is apparent from the Stokes stream function depicted in
the right half of Fig.~\ref{fig:numerical_snapshot}. For the pendant droplet, the flow at the interface is in the same direction as predicted by pure Marangoni flow in the absence of natural convection, i.e. from the apex towards the contact line. For the sessile droplet, however, the strong natural convection in the bulk transports water from the bulk to the apex. Thereby, despite of the enhanced evaporation rate at the top, there is a higher water concentration at the apex as compared to the contact line. This results in a Marangoni flow from the contact line towards the apex, i.e. in the opposite direction as predicted when considering the Marangoni effect without natural convection. Hence, the impact of the gravity is not only able to decisively control the flow direction in the bulk, but can also reverse the interfacial composition gradient and the corresponding Marangoni flow.

A quantitative comparison of the experimental and numerical results is
shown in Fig~\ref{fig:velocity_evolution}. Here, the temporal evolution
of the angular-averaged radial velocity of the micro-PIV measurement in
the plane close to the substrate is compared with the numerically
obtained radial velocity evaluated at the same height. It is apparent
that the radial velocity profile is in good agreement for the entire
drying time. For the case of the pendant droplet, however, the
simulation predicts an inward flow region close to the contact line,
which was not found in the corresponding micro-PIV measurement. This
difference can presumably be attributed to the assumption of
axisymmetry in the model, a condition that may easily be broken for the pendant
droplet system (cf. Fig~\ref{fig:exp_velocity}(b1)).


Gravitational effects triggered by density gradients due to selective evaporation can play a dominating role in controlling the flow in evaporating multicomponent droplets, even at the sub-millimeter scale and small Bond numbers. Misled by the small Bo, hitherto, most studies on this subject until now have disregarded the influence of natural convection. Our results show conclusively that natural convection can readily dominate the flow for $\mu$L droplets. Thus, our findings stimulate a careful treatment of the interplay of natural convection and Marangoni flow in multicomponent droplets in future studies. Furthermore, the possibility to reverse the bulk flow by overturning the system opens new application perspectives for surface coating and particle patterning. --On resubmission of our paper, we got aware of a reference which came to similar conclusion, however, employing different methods and different liquids~\cite{Edwards2018}.

This work is part of the FIP Industrial Partnership Programme (IPP) of the Netherlands Organization for Scientific Research (NWO). This research programme is co-financed by Oc\'{e}-Technologies B.V., University of Twente and Eindhoven University of Technology.




\begin{thebibliography}{43}%
\makeatletter
\providecommand \@ifxundefined [1]{%
 \@ifx{#1\undefined}
}%
\providecommand \@ifnum [1]{%
 \ifnum #1\expandafter \@firstoftwo
 \else \expandafter \@secondoftwo
 \fi
}%
\providecommand \@ifx [1]{%
 \ifx #1\expandafter \@firstoftwo
 \else \expandafter \@secondoftwo
 \fi
}%
\providecommand \natexlab [1]{#1}%
\providecommand \enquote  [1]{``#1''}%
\providecommand \bibnamefont  [1]{#1}%
\providecommand \bibfnamefont [1]{#1}%
\providecommand \citenamefont [1]{#1}%
\providecommand \href@noop [0]{\@secondoftwo}%
\providecommand \href [0]{\begingroup \@sanitize@url \@href}%
\providecommand \@href[1]{\@@startlink{#1}\@@href}%
\providecommand \@@href[1]{\endgroup#1\@@endlink}%
\providecommand \@sanitize@url [0]{\catcode `\\12\catcode `\$12\catcode
  `\&12\catcode `\#12\catcode `\^12\catcode `\_12\catcode `\%12\relax}%
\providecommand \@@startlink[1]{}%
\providecommand \@@endlink[0]{}%
\providecommand \url  [0]{\begingroup\@sanitize@url \@url }%
\providecommand \@url [1]{\endgroup\@href {#1}{\urlprefix }}%
\providecommand \urlprefix  [0]{URL }%
\providecommand \Eprint [0]{\href }%
\providecommand \doibase [0]{http://dx.doi.org/}%
\providecommand \selectlanguage [0]{\@gobble}%
\providecommand \bibinfo  [0]{\@secondoftwo}%
\providecommand \bibfield  [0]{\@secondoftwo}%
\providecommand \translation [1]{[#1]}%
\providecommand \BibitemOpen [0]{}%
\providecommand \bibitemStop [0]{}%
\providecommand \bibitemNoStop [0]{.\EOS\space}%
\providecommand \EOS [0]{\spacefactor3000\relax}%
\providecommand \BibitemShut  [1]{\csname bibitem#1\endcsname}%
\let\auto@bib@innerbib\@empty
\bibitem [{\citenamefont {Picknett}\ and\ \citenamefont
  {Bexon}(1977)}]{picknett1977}%
  \BibitemOpen
  \bibfield  {author} {\bibinfo {author} {\bibfnamefont {R.~G.}\ \bibnamefont
  {Picknett}}\ and\ \bibinfo {author} {\bibfnamefont {R.}~\bibnamefont
  {Bexon}},\ }\href@noop {} {\bibfield  {journal} {\bibinfo  {journal} {{J.
  Colloid Interface Sci.}}\ }\textbf {\bibinfo {volume} {{61}}},\ \bibinfo
  {pages} {336} (\bibinfo {year} {{1977}})}\BibitemShut {NoStop}%
\bibitem [{\citenamefont {Deegan}\ \emph {et~al.}(1997)\citenamefont {Deegan},
  \citenamefont {Bakajin}, \citenamefont {Dupont}, \citenamefont {Huber},
  \citenamefont {Nagel},\ and\ \citenamefont {Witten}}]{deegan1997}%
  \BibitemOpen
  \bibfield  {author} {\bibinfo {author} {\bibfnamefont {R.~D.}\ \bibnamefont
  {Deegan}}, \bibinfo {author} {\bibfnamefont {O.}~\bibnamefont {Bakajin}},
  \bibinfo {author} {\bibfnamefont {T.~F.}\ \bibnamefont {Dupont}}, \bibinfo
  {author} {\bibfnamefont {G.}~\bibnamefont {Huber}}, \bibinfo {author}
  {\bibfnamefont {S.~R.}\ \bibnamefont {Nagel}}, \ and\ \bibinfo {author}
  {\bibfnamefont {T.~A.}\ \bibnamefont {Witten}},\ }\href@noop {} {\bibfield
  {journal} {\bibinfo  {journal} {{Nature}}\ }\textbf {\bibinfo {volume}
  {{389}}},\ \bibinfo {pages} {827} (\bibinfo {year} {{1997}})}\BibitemShut
  {NoStop}%
\bibitem [{\citenamefont {Lohse}\ and\ \citenamefont
  {Zhang}(2015)}]{lohse2015rmp}%
  \BibitemOpen
  \bibfield  {author} {\bibinfo {author} {\bibfnamefont {D.}~\bibnamefont
  {Lohse}}\ and\ \bibinfo {author} {\bibfnamefont {X.}~\bibnamefont {Zhang}},\
  }\href@noop {} {\bibfield  {journal} {\bibinfo  {journal} {Rev. Mod. Phys.}\
  }\textbf {\bibinfo {volume} {87}},\ \bibinfo {pages} {981} (\bibinfo {year}
  {2015})}\BibitemShut {NoStop}%
\bibitem [{\citenamefont {Hu}\ and\ \citenamefont {Larson}(2002)}]{hu2002}%
  \BibitemOpen
  \bibfield  {author} {\bibinfo {author} {\bibfnamefont {H.}~\bibnamefont
  {Hu}}\ and\ \bibinfo {author} {\bibfnamefont {R.~G.}\ \bibnamefont
  {Larson}},\ }\href@noop {} {\bibfield  {journal} {\bibinfo  {journal} {{J.
  Phys. Chem. B.}}\ }\textbf {\bibinfo {volume} {{106}}},\ \bibinfo {pages}
  {1334} (\bibinfo {year} {{2002}})}\BibitemShut {NoStop}%
\bibitem [{\citenamefont {Popov}(2005)}]{popov2005}%
  \BibitemOpen
  \bibfield  {author} {\bibinfo {author} {\bibfnamefont {Y.~O.}\ \bibnamefont
  {Popov}},\ }\href@noop {} {\bibfield  {journal} {\bibinfo  {journal} {Phys.
  Rev. E.}\ }\textbf {\bibinfo {volume} {71}},\ \bibinfo {pages} {036313}
  (\bibinfo {year} {2005})}\BibitemShut {NoStop}%
\bibitem [{\citenamefont {Cazabat}\ and\ \citenamefont
  {Gu\'ena}(2010)}]{cazabat2010}%
  \BibitemOpen
  \bibfield  {author} {\bibinfo {author} {\bibfnamefont {A.~M.}\ \bibnamefont
  {Cazabat}}\ and\ \bibinfo {author} {\bibfnamefont {G.}~\bibnamefont
  {Gu\'ena}},\ }\href@noop {} {\bibfield  {journal} {\bibinfo  {journal} {Soft
  Matter}\ }\textbf {\bibinfo {volume} {6}},\ \bibinfo {pages} {2591} (\bibinfo
  {year} {2010})}\BibitemShut {NoStop}%
\bibitem [{\citenamefont {Shahidzadeh-Bonn}\ \emph {et~al.}(2006)\citenamefont
  {Shahidzadeh-Bonn}, \citenamefont {Rafai}, \citenamefont {Azouni},\ and\
  \citenamefont {Bonn}}]{sbonn2006}%
  \BibitemOpen
  \bibfield  {author} {\bibinfo {author} {\bibfnamefont {N.}~\bibnamefont
  {Shahidzadeh-Bonn}}, \bibinfo {author} {\bibfnamefont {S.}~\bibnamefont
  {Rafai}}, \bibinfo {author} {\bibfnamefont {A.}~\bibnamefont {Azouni}}, \
  and\ \bibinfo {author} {\bibfnamefont {D.}~\bibnamefont {Bonn}},\ }\href@noop
  {} {\bibfield  {journal} {\bibinfo  {journal} {{J. Fluid Mech.}}\ }\textbf
  {\bibinfo {volume} {{549}}},\ \bibinfo {pages} {307} (\bibinfo {year}
  {{2006}})}\BibitemShut {NoStop}%
\bibitem [{\citenamefont {Ristenpart}\ \emph {et~al.}(2007)\citenamefont
  {Ristenpart}, \citenamefont {Kim}, \citenamefont {Domingues}, \citenamefont
  {Wan},\ and\ \citenamefont {Stone}}]{ristenpart2007}%
  \BibitemOpen
  \bibfield  {author} {\bibinfo {author} {\bibfnamefont {W.~D.}\ \bibnamefont
  {Ristenpart}}, \bibinfo {author} {\bibfnamefont {P.~G.}\ \bibnamefont {Kim}},
  \bibinfo {author} {\bibfnamefont {C.}~\bibnamefont {Domingues}}, \bibinfo
  {author} {\bibfnamefont {J.}~\bibnamefont {Wan}}, \ and\ \bibinfo {author}
  {\bibfnamefont {H.~A.}\ \bibnamefont {Stone}},\ }\href@noop {} {\bibfield
  {journal} {\bibinfo  {journal} {{Phys. Rev. Lett.}}\ }\textbf {\bibinfo
  {volume} {{99}}},\ \bibinfo {pages} {234502} (\bibinfo {year}
  {{2007}})}\BibitemShut {NoStop}%
\bibitem [{\citenamefont {Sch\"onfeld}\ \emph {et~al.}(2008)\citenamefont
  {Sch\"onfeld}, \citenamefont {Graf}, \citenamefont {Hardt},\ and\
  \citenamefont {Butt}}]{schoenfeld2008}%
  \BibitemOpen
  \bibfield  {author} {\bibinfo {author} {\bibfnamefont {F.}~\bibnamefont
  {Sch\"onfeld}}, \bibinfo {author} {\bibfnamefont {K.~H.}\ \bibnamefont
  {Graf}}, \bibinfo {author} {\bibfnamefont {S.}~\bibnamefont {Hardt}}, \ and\
  \bibinfo {author} {\bibfnamefont {H.-J.}\ \bibnamefont {Butt}},\ }\href@noop
  {} {\bibfield  {journal} {\bibinfo  {journal} {Int. J. Heat Mass Transfer}\
  }\textbf {\bibinfo {volume} {51}},\ \bibinfo {pages} {3696} (\bibinfo {year}
  {2008})}\BibitemShut {NoStop}%
\bibitem [{\citenamefont {Gelderblom}\ \emph {et~al.}(2011)\citenamefont
  {Gelderblom}, \citenamefont {Marin}, \citenamefont {Nair}, \citenamefont {van
  Houselt}, \citenamefont {Lefferts}, \citenamefont {Snoeijer},\ and\
  \citenamefont {Lohse}}]{gelderblom2011}%
  \BibitemOpen
  \bibfield  {author} {\bibinfo {author} {\bibfnamefont {H.}~\bibnamefont
  {Gelderblom}}, \bibinfo {author} {\bibfnamefont {A.~G.}\ \bibnamefont
  {Marin}}, \bibinfo {author} {\bibfnamefont {H.}~\bibnamefont {Nair}},
  \bibinfo {author} {\bibfnamefont {A.}~\bibnamefont {van Houselt}}, \bibinfo
  {author} {\bibfnamefont {L.}~\bibnamefont {Lefferts}}, \bibinfo {author}
  {\bibfnamefont {J.~H.}\ \bibnamefont {Snoeijer}}, \ and\ \bibinfo {author}
  {\bibfnamefont {D.}~\bibnamefont {Lohse}},\ }\href@noop {} {\bibfield
  {journal} {\bibinfo  {journal} {Phys. Rev. E}\ }\textbf {\bibinfo {volume}
  {83}},\ \bibinfo {pages} {026306} (\bibinfo {year} {2011})}\BibitemShut
  {NoStop}%
\bibitem [{\citenamefont {Marin}\ \emph {et~al.}(2011)\citenamefont {Marin},
  \citenamefont {Gelderblom}, \citenamefont {Lohse},\ and\ \citenamefont
  {Snoeijer}}]{marin2011}%
  \BibitemOpen
  \bibfield  {author} {\bibinfo {author} {\bibfnamefont {A.~G.}\ \bibnamefont
  {Marin}}, \bibinfo {author} {\bibfnamefont {H.}~\bibnamefont {Gelderblom}},
  \bibinfo {author} {\bibfnamefont {D.}~\bibnamefont {Lohse}}, \ and\ \bibinfo
  {author} {\bibfnamefont {J.~H.}\ \bibnamefont {Snoeijer}},\ }\href@noop {}
  {\bibfield  {journal} {\bibinfo  {journal} {Phys. Rev. Lett.}\ }\textbf
  {\bibinfo {volume} {107}},\ \bibinfo {pages} {085502} (\bibinfo {year}
  {2011})}\BibitemShut {NoStop}%
\bibitem [{\citenamefont {Ledesma-Aguilar}\ \emph {et~al.}(2014)\citenamefont
  {Ledesma-Aguilar}, \citenamefont {Vella},\ and\ \citenamefont
  {Yeomans}}]{ledesma2014}%
  \BibitemOpen
  \bibfield  {author} {\bibinfo {author} {\bibfnamefont {R.}~\bibnamefont
  {Ledesma-Aguilar}}, \bibinfo {author} {\bibfnamefont {D.}~\bibnamefont
  {Vella}}, \ and\ \bibinfo {author} {\bibfnamefont {J.~M.}\ \bibnamefont
  {Yeomans}},\ }\href@noop {} {\bibfield  {journal} {\bibinfo  {journal} {Soft
  Matter}\ }\textbf {\bibinfo {volume} {10}},\ \bibinfo {pages} {8267}
  (\bibinfo {year} {2014})}\BibitemShut {NoStop}%
\bibitem [{\citenamefont {Tan}\ \emph {et~al.}(2016)\citenamefont {Tan},
  \citenamefont {Diddens}, \citenamefont {Lv}, \citenamefont {Kuerten},
  \citenamefont {Zhang},\ and\ \citenamefont {Lohse}}]{Tan2016}%
  \BibitemOpen
  \bibfield  {author} {\bibinfo {author} {\bibfnamefont {H.}~\bibnamefont
  {Tan}}, \bibinfo {author} {\bibfnamefont {C.}~\bibnamefont {Diddens}},
  \bibinfo {author} {\bibfnamefont {P.}~\bibnamefont {Lv}}, \bibinfo {author}
  {\bibfnamefont {J.~G.~M.}\ \bibnamefont {Kuerten}}, \bibinfo {author}
  {\bibfnamefont {X.}~\bibnamefont {Zhang}}, \ and\ \bibinfo {author}
  {\bibfnamefont {D.}~\bibnamefont {Lohse}},\ }\href@noop {} {\bibfield
  {journal} {\bibinfo  {journal} {Proc. Natl. Acad. Sci. U.S.A.}\ }\textbf
  {\bibinfo {volume} {113}},\ \bibinfo {pages} {8642} (\bibinfo {year}
  {2016})}\BibitemShut {NoStop}%
  \bibitem [{\citenamefont {Gatapova}\ \emph {et~al.}(2018)\citenamefont {Gatapova},
  \citenamefont {Shonina}, \citenamefont {Safonov}, \citenamefont {Sulyaeva},\ and\ \citenamefont
  {Kabov}}]{Gatapova2018}%
  \BibitemOpen
  \bibfield  {author} {\bibinfo {author} {\bibfnamefont {E. Y.}~\bibnamefont
  {Gatapova}}, \bibinfo {author} {\bibfnamefont {A. M.}~\bibnamefont {Shonina}}, \bibinfo
  {author} {\bibfnamefont {A. I.}~\bibnamefont {Safonov}}, \bibinfo {author}
  {\bibfnamefont {V. S.}~\bibnamefont {Sulyaeva}}, \ and\ \bibinfo {author} {\bibfnamefont
  {O. A.}~\bibnamefont {Kabov}},\ }\href@noop {} {\bibfield  {journal} {\bibinfo
  {journal} {Soft Matter}\ }\textbf {\bibinfo {volume} {14}},\ \bibinfo
  {pages} {1811} (\bibinfo {year} {2018})}\BibitemShut {NoStop}%
\bibitem [{\citenamefont {Brutin}\ \emph {et~al.}(2011)\citenamefont {Brutin},
  \citenamefont {Sobac}, \citenamefont {Loquet},\ and\ \citenamefont
  {Sampol}}]{brutin2011}%
  \BibitemOpen
  \bibfield  {author} {\bibinfo {author} {\bibfnamefont {D.}~\bibnamefont
  {Brutin}}, \bibinfo {author} {\bibfnamefont {B.}~\bibnamefont {Sobac}},
  \bibinfo {author} {\bibfnamefont {B.}~\bibnamefont {Loquet}}, \ and\ \bibinfo
  {author} {\bibfnamefont {J.}~\bibnamefont {Sampol}},\ }\href@noop {}
  {\bibfield  {journal} {\bibinfo  {journal} {{J. Fluid Mech.}}\ }\textbf
  {\bibinfo {volume} {{667}}},\ \bibinfo {pages} {85} (\bibinfo {year}
  {{2011}})}\BibitemShut {NoStop}%
\bibitem [{\citenamefont {Lim}\ \emph {et~al.}(2008)\citenamefont {Lim},
  \citenamefont {Lee}, \citenamefont {Lee}, \citenamefont {Lee}, \citenamefont
  {Park},\ and\ \citenamefont {Cho}}]{lim2008}%
  \BibitemOpen
  \bibfield  {author} {\bibinfo {author} {\bibfnamefont {J.~A.}\ \bibnamefont
  {Lim}}, \bibinfo {author} {\bibfnamefont {W.~H.}\ \bibnamefont {Lee}},
  \bibinfo {author} {\bibfnamefont {H.~S.}\ \bibnamefont {Lee}}, \bibinfo
  {author} {\bibfnamefont {J.~H.}\ \bibnamefont {Lee}}, \bibinfo {author}
  {\bibfnamefont {Y.~D.}\ \bibnamefont {Park}}, \ and\ \bibinfo {author}
  {\bibfnamefont {K.}~\bibnamefont {Cho}},\ }\href@noop {} {\bibfield
  {journal} {\bibinfo  {journal} {{Adv. Functional Mat.}}\ }\textbf {\bibinfo
  {volume} {{18}}},\ \bibinfo {pages} {229} (\bibinfo {year}
  {{2008}})}\BibitemShut {NoStop}%
\bibitem [{\citenamefont {Park}\ and\ \citenamefont
  {Moon}(2006)}]{park2006control}%
  \BibitemOpen
  \bibfield  {author} {\bibinfo {author} {\bibfnamefont {J.}~\bibnamefont
  {Park}}\ and\ \bibinfo {author} {\bibfnamefont {J.}~\bibnamefont {Moon}},\
  }\href@noop {} {\bibfield  {journal} {\bibinfo  {journal} {Langmuir}\
  }\textbf {\bibinfo {volume} {22}},\ \bibinfo {pages} {3506} (\bibinfo {year}
  {2006})}\BibitemShut {NoStop}%
\bibitem [{\citenamefont {Kong}\ \emph {et~al.}(2014)\citenamefont {Kong},
  \citenamefont {Tamargo}, \citenamefont {Kim}, \citenamefont {Johnson},
  \citenamefont {Gupta}, \citenamefont {Koh}, \citenamefont {Chin},
  \citenamefont {Steingart}, \citenamefont {Rand},\ and\ \citenamefont
  {McAlpine}}]{kong2014}%
  \BibitemOpen
  \bibfield  {author} {\bibinfo {author} {\bibfnamefont {Y.~L.}\ \bibnamefont
  {Kong}}, \bibinfo {author} {\bibfnamefont {I.~A.}\ \bibnamefont {Tamargo}},
  \bibinfo {author} {\bibfnamefont {H.}~\bibnamefont {Kim}}, \bibinfo {author}
  {\bibfnamefont {B.~N.}\ \bibnamefont {Johnson}}, \bibinfo {author}
  {\bibfnamefont {M.~K.}\ \bibnamefont {Gupta}}, \bibinfo {author}
  {\bibfnamefont {T.}~\bibnamefont {Koh}}, \bibinfo {author} {\bibfnamefont
  {H.}~\bibnamefont {Chin}}, \bibinfo {author} {\bibfnamefont {D.~A.}\
  \bibnamefont {Steingart}}, \bibinfo {author} {\bibfnamefont {B.~P.}\
  \bibnamefont {Rand}}, \ and\ \bibinfo {author} {\bibfnamefont {M.~C.}\
  \bibnamefont {McAlpine}},\ }\href@noop {} {\bibfield  {journal} {\bibinfo
  {journal} {Nano Letters}\ }\textbf {\bibinfo {volume} {14}},\ \bibinfo
  {pages} {7017} (\bibinfo {year} {2014})}\BibitemShut {NoStop}%
\bibitem [{\citenamefont {Jing}\ \emph {et~al.}(1998)\citenamefont {Jing},
  \citenamefont {Reed}, \citenamefont {Huang}, \citenamefont {Hu},
  \citenamefont {Clarke}, \citenamefont {Edington}, \citenamefont {Housman},
  \citenamefont {Anantharaman}, \citenamefont {Huff},\ and\ \citenamefont
  {Mishra}}]{jing1998automated}%
  \BibitemOpen
  \bibfield  {author} {\bibinfo {author} {\bibfnamefont {J.}~\bibnamefont
  {Jing}}, \bibinfo {author} {\bibfnamefont {J.}~\bibnamefont {Reed}}, \bibinfo
  {author} {\bibfnamefont {J.}~\bibnamefont {Huang}}, \bibinfo {author}
  {\bibfnamefont {X.}~\bibnamefont {Hu}}, \bibinfo {author} {\bibfnamefont
  {V.}~\bibnamefont {Clarke}}, \bibinfo {author} {\bibfnamefont
  {J.}~\bibnamefont {Edington}}, \bibinfo {author} {\bibfnamefont
  {D.}~\bibnamefont {Housman}}, \bibinfo {author} {\bibfnamefont {T.~S.}\
  \bibnamefont {Anantharaman}}, \bibinfo {author} {\bibfnamefont {E.~J.}\
  \bibnamefont {Huff}}, \ and\ \bibinfo {author} {\bibfnamefont
  {B.}~\bibnamefont {Mishra}},\ }\href@noop {} {\bibfield  {journal} {\bibinfo
  {journal} {Proc. Natl. Acad. Sci. U.S.A.}\ }\textbf {\bibinfo {volume}
  {95}},\ \bibinfo {pages} {8046} (\bibinfo {year} {1998})}\BibitemShut
  {NoStop}%
\bibitem [{\citenamefont {Hu}\ and\ \citenamefont {Larson}(2006)}]{hu2006}%
  \BibitemOpen
  \bibfield  {author} {\bibinfo {author} {\bibfnamefont {H.}~\bibnamefont
  {Hu}}\ and\ \bibinfo {author} {\bibfnamefont {R.~G.}\ \bibnamefont
  {Larson}},\ }\href@noop {} {\bibfield  {journal} {\bibinfo  {journal} {J.
  Phys. Chem. B}\ }\textbf {\bibinfo {volume} {110}},\ \bibinfo {pages} {7090}
  (\bibinfo {year} {2006})}\BibitemShut {NoStop}%
\bibitem [{\citenamefont {Tam}\ \emph {et~al.}(2009)\citenamefont {Tam},
  \citenamefont {von Arnim}, \citenamefont {McKinley},\ and\ \citenamefont
  {Hosoi}}]{tam2009}%
  \BibitemOpen
  \bibfield  {author} {\bibinfo {author} {\bibfnamefont {D.}~\bibnamefont
  {Tam}}, \bibinfo {author} {\bibfnamefont {V.}~\bibnamefont {von Arnim}},
  \bibinfo {author} {\bibfnamefont {G.~H.}\ \bibnamefont {McKinley}}, \ and\
  \bibinfo {author} {\bibfnamefont {A.~E.}\ \bibnamefont {Hosoi}},\ }\href@noop
  {} {\bibfield  {journal} {\bibinfo  {journal} {J. Fluid Mech.}\ }\textbf
  {\bibinfo {volume} {624}},\ \bibinfo {pages} {101} (\bibinfo {year}
  {2009})}\BibitemShut {NoStop}%
\bibitem [{\citenamefont {Christy}\ \emph {et~al.}(2011)\citenamefont
  {Christy}, \citenamefont {Hamamoto},\ and\ \citenamefont
  {Sefiane}}]{christy2011}%
  \BibitemOpen
  \bibfield  {author} {\bibinfo {author} {\bibfnamefont {J.~R.~E.}\
  \bibnamefont {Christy}}, \bibinfo {author} {\bibfnamefont {Y.}~\bibnamefont
  {Hamamoto}}, \ and\ \bibinfo {author} {\bibfnamefont {K.}~\bibnamefont
  {Sefiane}},\ }\href@noop {} {\bibfield  {journal} {\bibinfo  {journal} {Phys.
  Rev. Lett.}\ }\textbf {\bibinfo {volume} {106}},\ \bibinfo {pages} {205701}
  (\bibinfo {year} {2011})}\BibitemShut {NoStop}%
\bibitem [{\citenamefont {Bennacer}\ and\ \citenamefont
  {Sefiane}(2014)}]{bennacer_sefiane_2014}%
  \BibitemOpen
  \bibfield  {author} {\bibinfo {author} {\bibfnamefont {R.}~\bibnamefont
  {Bennacer}}\ and\ \bibinfo {author} {\bibfnamefont {K.}~\bibnamefont
  {Sefiane}},\ }\href@noop {} {\bibfield  {journal} {\bibinfo  {journal} {J.
  Fluid Mech.}\ }\textbf {\bibinfo {volume} {749}},\ \bibinfo {pages} {649}
  (\bibinfo {year} {2014})}\BibitemShut {NoStop}%
\bibitem [{\citenamefont {Tan}\ \emph {et~al.}(2017)\citenamefont {Tan},
  \citenamefont {Diddens}, \citenamefont {Versluis}, \citenamefont {Butt},
  \citenamefont {Lohse},\ and\ \citenamefont {Zhang}}]{Tan2017}%
  \BibitemOpen
  \bibfield  {author} {\bibinfo {author} {\bibfnamefont {H.}~\bibnamefont
  {Tan}}, \bibinfo {author} {\bibfnamefont {C.}~\bibnamefont {Diddens}},
  \bibinfo {author} {\bibfnamefont {M.}~\bibnamefont {Versluis}}, \bibinfo
  {author} {\bibfnamefont {H.-J.}\ \bibnamefont {Butt}}, \bibinfo {author}
  {\bibfnamefont {D.}~\bibnamefont {Lohse}}, \ and\ \bibinfo {author}
  {\bibfnamefont {X.}~\bibnamefont {Zhang}},\ }\href@noop {} {\bibfield
  {journal} {\bibinfo  {journal} {Soft Matter}\ }\textbf {\bibinfo {volume}
  {13}},\ \bibinfo {pages} {2749} (\bibinfo {year} {2017})}\BibitemShut
  {NoStop}%
\bibitem [{\citenamefont {Li}\ \emph {et~al.}(2018)\citenamefont {Li},
  \citenamefont {Lv}, \citenamefont {Diddens}, \citenamefont {Tan},
  \citenamefont {Wijshoff}, \citenamefont {Versluis},\ and\ \citenamefont
  {Lohse}}]{Li2018}%
  \BibitemOpen
  \bibfield  {author} {\bibinfo {author} {\bibfnamefont {Y.}~\bibnamefont
  {Li}}, \bibinfo {author} {\bibfnamefont {P.}~\bibnamefont {Lv}}, \bibinfo
  {author} {\bibfnamefont {C.}~\bibnamefont {Diddens}}, \bibinfo {author}
  {\bibfnamefont {H.}~\bibnamefont {Tan}}, \bibinfo {author} {\bibfnamefont
  {H.}~\bibnamefont {Wijshoff}}, \bibinfo {author} {\bibfnamefont
  {M.}~\bibnamefont {Versluis}}, \ and\ \bibinfo {author} {\bibfnamefont
  {D.}~\bibnamefont {Lohse}},\ }\href@noop {} {\bibfield  {journal} {\bibinfo
  {journal} {Phys. Rev. Lett.}\ }\textbf {\bibinfo {volume} {120}},\ \bibinfo
  {pages} {224501} (\bibinfo {year} {2018})}\BibitemShut {NoStop}%
\bibitem [{\citenamefont {Kim}\ and\ \citenamefont {Stone}(2018)}]{Kim2018}%
  \BibitemOpen
  \bibfield  {author} {\bibinfo {author} {\bibfnamefont {H.}~\bibnamefont
  {Kim}}\ and\ \bibinfo {author} {\bibfnamefont {H.}~\bibnamefont {Stone}},\
  }\href@noop {} {\bibfield  {journal} {\bibinfo  {journal} {J. Fluid Mech.}\
  }\textbf {\bibinfo {volume} {850}},\ \bibinfo {pages} {769} (\bibinfo {year}
  {2018})}\BibitemShut {NoStop}%
\bibitem [{\citenamefont {Kim}\ \emph {et~al.}(2016)\citenamefont {Kim},
  \citenamefont {Boulogne}, \citenamefont {Um}, \citenamefont {Jacobi},
  \citenamefont {Button},\ and\ \citenamefont {Stone}}]{kim2016controlled}%
  \BibitemOpen
  \bibfield  {author} {\bibinfo {author} {\bibfnamefont {H.}~\bibnamefont
  {Kim}}, \bibinfo {author} {\bibfnamefont {F.}~\bibnamefont {Boulogne}},
  \bibinfo {author} {\bibfnamefont {E.}~\bibnamefont {Um}}, \bibinfo {author}
  {\bibfnamefont {I.}~\bibnamefont {Jacobi}}, \bibinfo {author} {\bibfnamefont
  {E.}~\bibnamefont {Button}}, \ and\ \bibinfo {author} {\bibfnamefont {H.~A.}\
  \bibnamefont {Stone}},\ }\href@noop {} {\bibfield  {journal} {\bibinfo
  {journal} {Phys. Rev. Lett.}\ }\textbf {\bibinfo {volume} {116}},\ \bibinfo
  {pages} {124501} (\bibinfo {year} {2016})}\BibitemShut {NoStop}%
\bibitem [{\citenamefont {Shin}\ \emph {et~al.}(2009)\citenamefont {Shin},
  \citenamefont {Lee}, \citenamefont {Jung},\ and\ \citenamefont
  {Y.}}]{SHIN2009}%
  \BibitemOpen
  \bibfield  {author} {\bibinfo {author} {\bibfnamefont {D.~H.}\ \bibnamefont
  {Shin}}, \bibinfo {author} {\bibfnamefont {S.~H.}\ \bibnamefont {Lee}},
  \bibinfo {author} {\bibfnamefont {J.-Y.}\ \bibnamefont {Jung}}, \ and\
  \bibinfo {author} {\bibfnamefont {Y.~J.}\ \bibnamefont {Y.}},\ }\href@noop {}
  {\bibfield  {journal} {\bibinfo  {journal} {Microelectron. Eng.}\ }\textbf
  {\bibinfo {volume} {86}},\ \bibinfo {pages} {1350 } (\bibinfo {year}
  {2009})}\BibitemShut {NoStop}%
\bibitem [{\citenamefont {Savino}\ and\ \citenamefont
  {Monti}(1996)}]{SAVINO1996}%
  \BibitemOpen
  \bibfield  {author} {\bibinfo {author} {\bibfnamefont {R.}~\bibnamefont
  {Savino}}\ and\ \bibinfo {author} {\bibfnamefont {R.}~\bibnamefont {Monti}},\
  }\href@noop {} {\bibfield  {journal} {\bibinfo  {journal} {J. Cryst. Growth}\
  }\textbf {\bibinfo {volume} {165}},\ \bibinfo {pages} {308 } (\bibinfo {year}
  {1996})}\BibitemShut {NoStop}%
\bibitem [{\citenamefont {Kang}\ \emph {et~al.}(2013)\citenamefont {Kang},
  \citenamefont {Lim}, \citenamefont {Lee},\ and\ \citenamefont
  {Lee}}]{kang2013}%
  \BibitemOpen
  \bibfield  {author} {\bibinfo {author} {\bibfnamefont {K.~H.}\ \bibnamefont
  {Kang}}, \bibinfo {author} {\bibfnamefont {H.~C.}\ \bibnamefont {Lim}},
  \bibinfo {author} {\bibfnamefont {H.~W.}\ \bibnamefont {Lee}}, \ and\
  \bibinfo {author} {\bibfnamefont {S.~J.}\ \bibnamefont {Lee}},\ }\href@noop
  {} {\bibfield  {journal} {\bibinfo  {journal} {Phys. Fluids}\ }\textbf
  {\bibinfo {volume} {25}},\ \bibinfo {pages} {042001} (\bibinfo {year}
  {2013})}\BibitemShut {NoStop}%
\bibitem [{\citenamefont {Pradhan}\ and\ \citenamefont
  {Panigrahi}(2016)}]{kumar2016}%
  \BibitemOpen
  \bibfield  {author} {\bibinfo {author} {\bibfnamefont {T.~K.}\ \bibnamefont
  {Pradhan}}\ and\ \bibinfo {author} {\bibfnamefont {P.~K.}\ \bibnamefont
  {Panigrahi}},\ }\href@noop {} {\bibfield  {journal} {\bibinfo  {journal}
  {Colloids Surf. A}\ }\textbf {\bibinfo {volume} {500}},\ \bibinfo {pages}
  {154 } (\bibinfo {year} {2016})}\BibitemShut {NoStop}%
\bibitem [{\citenamefont {Pradhan}\ and\ \citenamefont
  {Panigrahi}(2017)}]{kumar2017}%
  \BibitemOpen
  \bibfield  {author} {\bibinfo {author} {\bibfnamefont {T.~K.}\ \bibnamefont
  {Pradhan}}\ and\ \bibinfo {author} {\bibfnamefont {P.~K.}\ \bibnamefont
  {Panigrahi}},\ }\href@noop {} {\bibfield  {journal} {\bibinfo  {journal}
  {Colloids Surf. A}\ }\textbf {\bibinfo {volume} {530}},\ \bibinfo {pages} {1
  } (\bibinfo {year} {2017})}\BibitemShut {NoStop}%
\bibitem [{\citenamefont {Kumar}\ and\ \citenamefont
  {Mandal}(2018)}]{KUMAR2018}%
  \BibitemOpen
  \bibfield  {author} {\bibinfo {author} {\bibfnamefont {A.}~\bibnamefont
  {Kumar}}\ and\ \bibinfo {author} {\bibfnamefont {D.~K.}\ \bibnamefont
  {Mandal}},\ }\href@noop {} {\bibfield  {journal} {\bibinfo  {journal}
  {International Journal of Multiphase Flow}\ }\textbf {\bibinfo {volume}
  {102}},\ \bibinfo {pages} {130 } (\bibinfo {year} {2018})}\BibitemShut
  {NoStop}%
\bibitem [{soa(1990)}]{soap1990}%
  \BibitemOpen
  \href@noop {} {\bibfield  {journal} {\bibinfo  {journal} {Technical Report of
  the Soap and Detergent Association}\ ,\ \bibinfo {pages} {212}} (\bibinfo
  {year} {1990})}\BibitemShut {NoStop}%
\bibitem [{\citenamefont {Sousa}(1995)}]{Sousa1995}%
  \BibitemOpen
  \bibfield  {author} {\bibinfo {author} {\bibfnamefont {R.}~\bibnamefont
  {Sousa}},\ }\href@noop {} {\bibfield  {journal} {\bibinfo  {journal} {Acta
  Cryst. D}\ }\textbf {\bibinfo {volume} {51}},\ \bibinfo {pages} {271}
  (\bibinfo {year} {1995})}\BibitemShut {NoStop}%
\bibitem [{gly(1963)}]{gly1963}%
  \BibitemOpen
  \href@noop {} {\bibfield  {journal} {\bibinfo  {journal} {Glycerine Producers
  Association, New York}\ } (\bibinfo {year} {1963})}\BibitemShut {NoStop}%
\bibitem [{\citenamefont {Shin}\ \emph {et~al.}(2016)\citenamefont {Shin},
  \citenamefont {Jacobi},\ and\ \citenamefont {Stone}}]{shin2016}%
  \BibitemOpen
  \bibfield  {author} {\bibinfo {author} {\bibfnamefont {S.}~\bibnamefont
  {Shin}}, \bibinfo {author} {\bibfnamefont {I.}~\bibnamefont {Jacobi}}, \ and\
  \bibinfo {author} {\bibfnamefont {H.~A.}\ \bibnamefont {Stone}},\ }\href@noop
  {} {\bibfield  {journal} {\bibinfo  {journal} {EPL}\ }\textbf {\bibinfo
  {volume} {113}},\ \bibinfo {pages} {24002} (\bibinfo {year}
  {2016})}\BibitemShut {NoStop}%
  \bibitem {SM}
  {See Supplemental Material at [url] for further details on experimental and numerical methods. The supplemental Materials includes also evidences on arguments about Archimedes number from experiments and numerical simulations. The supplemental Material includes Refs. [39-43,48,50]}\BibitemShut {NoStop}%
\bibitem [{\citenamefont {Thielicke}\ and\ \citenamefont
  {Stamhuis}(2014)}]{pivlab2014}%
  \BibitemOpen
  \bibfield  {author} {\bibinfo {author} {\bibfnamefont {W.}~\bibnamefont
  {Thielicke}}\ and\ \bibinfo {author} {\bibfnamefont {E.~J.}\ \bibnamefont
  {Stamhuis}},\ } {\bibfield
  {journal} {\bibinfo  {journal} {Journal of Open Research Software}\ }
  (\bibinfo {year} {2014})}\BibitemShut
  {NoStop}%
\bibitem [{\citenamefont {Garcia}(2011)}]{pivsmooth2011}%
  \BibitemOpen
  \bibfield  {author} {\bibinfo {author} {\bibfnamefont {D.}~\bibnamefont
  {Garcia}},\ } {\bibfield  {journal}
  {\bibinfo  {journal} {Exp. Fluids}\ }\textbf {\bibinfo {volume} {50}},\
  \bibinfo {pages} {1247} (\bibinfo {year} {2011})}\BibitemShut {NoStop}%
\bibitem [{\citenamefont {Peng}\ \emph {et~al.}(2014)\citenamefont {Peng},
  \citenamefont {Xu}, \citenamefont {Hughes},\ and\ \citenamefont
  {Zhang}}]{Peng2014}%
  \BibitemOpen
  \bibfield  {author} {\bibinfo {author} {\bibfnamefont {S.}~\bibnamefont
  {Peng}}, \bibinfo {author} {\bibfnamefont {C.}~\bibnamefont {Xu}}, \bibinfo
  {author} {\bibfnamefont {T.~C.}\ \bibnamefont {Hughes}}, \ and\ \bibinfo
  {author} {\bibfnamefont {X.}~\bibnamefont {Zhang}},\ }\href@noop {}
  {\bibfield  {journal} {\bibinfo  {journal} {Langmuir}\ }\textbf {\bibinfo
  {volume} {30}},\ \bibinfo {pages} {12270} (\bibinfo {year}
  {2014})}\BibitemShut {NoStop}%
\bibitem [{\citenamefont {Raffel}(2018)}]{Raffel2018}%
  \BibitemOpen
  \bibfield  {author} {\bibinfo {author} {\bibfnamefont {M.}~\bibnamefont
  {Raffel}},\bibinfo {author} {\bibfnamefont {C.~E.}~\bibnamefont {Willert}}, \bibinfo {author}
  {\bibfnamefont {F.}~\bibnamefont {Scarano}}, \bibinfo {author}
  {\bibfnamefont {C.~J.}\ \bibnamefont {K$\ddot{\text{a}}$hler}}, \bibinfo {author}
  {\bibfnamefont {S.~T.}~\bibnamefont {Wereley}}, \ and\ \bibinfo {author}
  {\bibfnamefont {J.}~\bibnamefont {Kompenhans}},\ } {\bibfield
  {journal} {\bibinfo  {journal} {Particle Image Velocimetry}\ },\ \bibinfo {pages} {Third Edition} (\bibinfo {year}
  {2018})}\BibitemShut {NoStop}%
  \bibitem [{\citenamefont {Diddens}(2017)}]{Diddens2017b}%
  \BibitemOpen
  \bibfield  {author} {\bibinfo {author} {\bibfnamefont {C.}~\bibnamefont
  {Diddens}},\ }\href@noop {} {\bibfield  {journal} {\bibinfo  {journal} {J.
  Comput. Phys.}\ }\textbf {\bibinfo {volume} {340}},\ \bibinfo {pages} {670}
  (\bibinfo {year} {2017})}\BibitemShut {NoStop}%
\bibitem [{\citenamefont {Xue}\ \emph {et~al.}(2017)\citenamefont {Xue},
  \citenamefont {Khodaparast}, \citenamefont {Zhu}, \citenamefont {Nunes},
  \citenamefont {Kim},\ and\ \citenamefont {Stone}}]{nan2017}%
  \BibitemOpen
  \bibfield  {author} {\bibinfo {author} {\bibfnamefont {N.}~\bibnamefont
  {Xue}}, \bibinfo {author} {\bibfnamefont {S.}~\bibnamefont {Khodaparast}},
  \bibinfo {author} {\bibfnamefont {L.}~\bibnamefont {Zhu}}, \bibinfo {author}
  {\bibfnamefont {J.~K.}\ \bibnamefont {Nunes}}, \bibinfo {author}
  {\bibfnamefont {H.}~\bibnamefont {Kim}}, \ and\ \bibinfo {author}
  {\bibfnamefont {H.~A.}\ \bibnamefont {Stone}},\ }\href@noop {} {\bibfield
  {journal} {\bibinfo  {journal} {Nat. Commun.}\ }\textbf {\bibinfo {volume}
  {8}},\ \bibinfo {pages} {1960} (\bibinfo {year} {2017})}\BibitemShut
  {NoStop}%
\bibitem [{\citenamefont {Yu}\ \emph {et~al.}(2015)\citenamefont {Yu},
  \citenamefont {Lu}, \citenamefont {Lohse},\ and\ \citenamefont
  {Zhang}}]{yu2015}%
  \BibitemOpen
  \bibfield  {author} {\bibinfo {author} {\bibfnamefont {H.}~\bibnamefont
  {Yu}}, \bibinfo {author} {\bibfnamefont {Z.}~\bibnamefont {Lu}}, \bibinfo
  {author} {\bibfnamefont {D.}~\bibnamefont {Lohse}}, \ and\ \bibinfo {author}
  {\bibfnamefont {X.}~\bibnamefont {Zhang}},\ }\href@noop {} {\bibfield
  {journal} {\bibinfo  {journal} {Langmuir}\ }\textbf {\bibinfo {volume}
  {31}},\ \bibinfo {pages} {12628} (\bibinfo {year} {2015})}\BibitemShut
  {NoStop}%
\bibitem [{\citenamefont {Tanaka}\ \emph {et~al.}(1988)\citenamefont {Tanaka},
  \citenamefont {Ohta}, \citenamefont {Kubota},\ and\ \citenamefont
  {Makita}}]{Tanaka1988}%
  \BibitemOpen
  \bibfield  {author} {\bibinfo {author} {\bibfnamefont {Y.}~\bibnamefont
  {Tanaka}}, \bibinfo {author} {\bibfnamefont {K.}~\bibnamefont {Ohta}},
  \bibinfo {author} {\bibfnamefont {H.}~\bibnamefont {Kubota}}, \ and\ \bibinfo
  {author} {\bibfnamefont {T.}~\bibnamefont {Makita}},\ }\href@noop {}
  {\bibfield  {journal} {\bibinfo  {journal} {Int. J. Thermophys}\ }\textbf
  {\bibinfo {volume} {9}},\ \bibinfo {pages} {511} (\bibinfo {year}
  {1988})}\BibitemShut {NoStop}%
  \bibitem [{\citenamefont {Nakanishi}\ \emph {et~al.}(1971)\citenamefont {Nakanishi},
  \citenamefont {Matsumoto}, \ and\ \citenamefont
  {Hayatsu}}]{Nakanishi1971}%
  \BibitemOpen
  \bibfield  {author} {\bibinfo {author} {\bibfnamefont {K.}~\bibnamefont
  {Nakanishi}}, \bibinfo {author} {\bibfnamefont {T.}~\bibnamefont {Matsumoto}}, \ and\ \bibinfo {author}
  {\bibfnamefont {M.}~\bibnamefont {Hayatsu}},\ }\href@noop {} {\bibfield
  {journal} {\bibinfo  {journal} {J. Chem. Eng. Data}\ }\textbf {\bibinfo {volume}
  {16}},\ \bibinfo {pages} {44} (\bibinfo {year} {1971})}\BibitemShut
  {NoStop}%
  \bibitem [{\citenamefont {Diddens}\ \emph
  {et~al.}(2017{\natexlab{a}})\citenamefont {Diddens}, \citenamefont {Tan},
  \citenamefont {Lv}, \citenamefont {Versluis}, \citenamefont {Kuerten},
  \citenamefont {Zhang},\ and\ \citenamefont {Lohse}}]{Diddens2017c}%
  \BibitemOpen
  \bibfield  {author} {\bibinfo {author} {\bibfnamefont {C.}~\bibnamefont
  {Diddens}}, \bibinfo {author} {\bibfnamefont {H.}~\bibnamefont {Tan}},
  \bibinfo {author} {\bibfnamefont {P.}~\bibnamefont {Lv}}, \bibinfo {author}
  {\bibfnamefont {M.}~\bibnamefont {Versluis}}, \bibinfo {author}
  {\bibfnamefont {J.}~\bibnamefont {Kuerten}}, \bibinfo {author} {\bibfnamefont
  {X.}~\bibnamefont {Zhang}}, \ and\ \bibinfo {author} {\bibfnamefont
  {D.}~\bibnamefont {Lohse}},\ }\href@noop {} {\bibfield  {journal} {\bibinfo
  {journal} {J. Fluid Mech.}\ }\textbf {\bibinfo {volume} {823}},\ \bibinfo
  {pages} {470} (\bibinfo {year} {2017}{\natexlab{a}})}\BibitemShut {NoStop}%
\bibitem [{\citenamefont {Marcolli}\ and\ \citenamefont
  {Peter}(2005)}]{marcolli2005}%
  \BibitemOpen
  \bibfield  {author} {\bibinfo {author} {\bibfnamefont {C.}~\bibnamefont
  {Marcolli}}\ and\ \bibinfo {author} {\bibfnamefont {T.}~\bibnamefont
  {Peter}},\ }\href@noop {} {\bibfield  {journal} {\bibinfo  {journal} {Atmos.
  Chem. Phys.}\ }\textbf {\bibinfo {volume} {5}},\ \bibinfo {pages} {1545}
  (\bibinfo {year} {2005})}\BibitemShut {NoStop}%
\bibitem [{\citenamefont {Diddens}\ \emph
  {et~al.}(2017{\natexlab{b}})\citenamefont {Diddens}, \citenamefont {Kuerten},
  \citenamefont {van~der Geld},\ and\ \citenamefont {Wijshoff}}]{Diddens2017a}%
  \BibitemOpen
  \bibfield  {author} {\bibinfo {author} {\bibfnamefont {C.}~\bibnamefont
  {Diddens}}, \bibinfo {author} {\bibfnamefont {J.}~\bibnamefont {Kuerten}},
  \bibinfo {author} {\bibfnamefont {C.}~\bibnamefont {van~der Geld}}, \ and\
  \bibinfo {author} {\bibfnamefont {H.}~\bibnamefont {Wijshoff}},\ }\href@noop
  {} {\bibfield  {journal} {\bibinfo  {journal} {J. Colloid Interf. Sci.}\
  }\textbf {\bibinfo {volume} {487}},\ \bibinfo {pages} {426} (\bibinfo {year}
  {2017}{\natexlab{b}})}\BibitemShut {NoStop}%
\bibitem [{\citenamefont {Edwards}\ and\ \citenamefont
  {Atkinson}\ and\ \citenamefont {Cheung}\  and\ \citenamefont {Liang}\ and\ \citenamefont {Fairhurst}\ \citenamefont {Ouali}(2018)}]{Edwards2018}%
  \BibitemOpen
  \bibfield  {author} {\bibinfo {author} {\bibfnamefont {A.}~\bibnamefont
  {M.}~\bibnamefont
  {J.}~\bibnamefont
  {Edwards}},\  \bibinfo {author} {\bibfnamefont {P.}~\bibnamefont
  {S.}~\bibnamefont
  {Atkinson}},\ \bibinfo {author} {\bibfnamefont {C.}~\bibnamefont
  {S.}~\bibnamefont
  {Cheung}},\ \bibinfo {author} {\bibfnamefont {H.}~\bibnamefont
  {Liang}},\ \bibinfo {author} {\bibfnamefont {D.}~\bibnamefont
  {J.}~\bibnamefont
  {Fairhurst}}, and \bibinfo {author} {\bibfnamefont {F.}~\bibnamefont
  {F.}~\bibnamefont
  {Ouali}},\ }\href@noop {} {\bibfield  {journal} {\bibinfo  {journal} {Phys. Rev. Lett}\ }\textbf {\bibinfo {volume} {121}},\ \bibinfo {pages} {184501}
  (\bibinfo {year} {2018})}\BibitemShut {NoStop}%
\end{thebibliography}

%


\end{document}